\documentclass[preprint]{aastex}
\usepackage{amsfonts}
\usepackage{subfigure}
\usepackage{color}

\usepackage{natbib}

\newcommand{\h}{\ensuremath{\, {\rm h}}}

\newcommand{\cm}{\ensuremath{\,{\rm cm}}}
\newcommand{\m}{\ensuremath{\,{\rm m}}}

\newcommand{\Mpc}{\ensuremath{\,{\rm Mpc}}}
\newcommand{\K}{\ensuremath{\, {\rm K}}}
\newcommand{\mK}{\ensuremath{\, {\rm mK}}}

\newcommand{\MHz}{\ensuremath{\, {\rm MHz}}}

\newcommand{\Msun}{\ensuremath{{M_\sun}}}

\newcommand{\bfC}{\mathbf{C}}

\begin{document}

\title{Forecasts on the Dark Energy and Primordial Non-Gaussianity Observations
 with the Tianlai Cylinder Array}

\author{{\large Yidong Xu\altaffilmark{1}, Xin Wang\altaffilmark{2},
Xuelei Chen\altaffilmark{1,3}
\\}}

\altaffiltext{1}{National Astronomical Observatories, Chinese Academy of Sciences, Beijing 100012, China}
\altaffiltext{2}{Department of Physics \& Astronomy, Johns Hopkins University,
Baltimore, MD, US, 21218}
\altaffiltext{3}{Center for High Energy Physics, Peking University, Beijing 100871, China}


\begin{abstract}
The Tianlai experiment is dedicated to the observation of large
scale structures (LSS) by the 21 cm intensity mapping technique.
In this paper we make forecasts on its capability at observing or constraining the
dark energy parameters and the primordial non-Gaussianity. From the LSS data one can
use the baryon acoustic oscillation (BAO) and the growth rate derived from
the redshift space distortion (RSD) to measure the dark energy density
and equation of state. The primordial non-Gaussianity can be constrained either
by looking for scale-dependent bias in the power spectrum, or by using the
bispectrum. Here we consider three cases: the Tianlai cylinder array pathfinder which is
currently being built, an upgrade of the pathfinder array with more
receiver units, and the full-scale Tianlai cylinder array.
Using the full-scale Tianlai experiment, we expect $\sigma_{w_0} \sim 0.082$ and $\sigma_{w_a} \sim 0.21$
from the BAO and RSD measurements, $\sigma_{\rm f_{NL}}^{\rm local} \sim 14$ from
the power spectrum measurements with scale-dependent bias, and
$\sigma_{\rm f_{NL}}^{\rm local} \sim 22$ and $\sigma_{\rm f_{NL}}^{\rm equil} \sim 157$
from the bispectrum measurements.
\end{abstract}



\section{Introduction}\label{Intro}
Baryon acoustic oscillations (BAO) are the frozen sound waves which were present in the photon-baryon plasma prior
to the recombination epoch, and they imprint features on the cosmic microware background (CMB)
as well as the large scale structures in the later Universe.
The (comoving) characteristic scale of the BAO is determined by the sound horizon at the
last scattering surface:
\begin{eqnarray}
  s &=& \int_0^{t_{\rm rec}}\,c_s\, (1+z)dt
     = \int_{z_{\rm rec}}^\infty {c_s\;dz\over H(z)} ,
\label{eq:rsound}
\end{eqnarray}
where $c_s$ is the sound speed, $H(z)$ is the Hubble expansion rate,
and $t_{\rm rec}$ and $z_{\rm rec}$ are the recombination time and redshift respectively.
The BAO scale provides a standard ruler to
measure the angular diameter distance $D_A(z)$ and the Hubble
parameter $H(z)$ \citep{2003ApJ...598..720S,2003ApJ...594..665B}, hence serves as a promising tool to constrain the properties of dark energy,
which determines the expansion rate of the Universe.
This technique has successfully been used to put cosmological constraints on
dark energy parameters from optical surveys (e.g. \citealt{2014MNRAS.441...24A}).

In addition to observing the cosmic expansion history through 
the BAO features in the matter power spectrum,
a large scale structure measurement can also yield the structure growth rate
$f(z)$ from redshift space distortions (RSD). 
The RSD features in the galaxy distribution have also been used to
constrain the cosmological parameters, and to
distinguish dark energy and various modified gravity models
(e.g. \citealt{2008Natur.451..541G,2008JCAP...05..021W,2008APh....29..336L,
2014MNRAS.443.1065B,2014MNRAS.439.3504S}).
The BAO and RSD features complement each other, and may also help break the 
degeneracy in dark energy and modified gravity models.

The large scale structures in the Universe could also be measured through the 
21 cm emission from neutral hydrogen (HI) in galaxies,
though current radio observations of HI in galaxies,
e.g. the ALFALFA survey \citep{2005AJ....130.2598G}
have been limited to redshift $z \lesssim 0.2$ at present. While some
telescopes which are currently being constructed, e.g. the
FAST \citep{2000ASPC..213..523N},
ASKAP \citep{2008ExA....22..151J,2009arXiv0910.2935B}, and
MeerKAT\citep{2000ASPC..213..523N}, will have much greater sensitivities,
HI survey of galaxies at high redshift would still be a challenging task.
Even for the Square Kilometer Array (SKA), with its wide area HI
galaxy survey, the constraining power on cosmological parameters is only
comparable to the existing optical galaxy surveys
\citep{2004NewAR..48.1013R,2010MNRAS.401..743A}.

However, power spectrum constraints on cosmological paramters can be efficiently
obtained with dedicated telescopes with lower resolution than galaxy surveys.
Instead, one could observe in the {\it intensity mapping}
mode, in which each pixel or voxel contains many galaxies. Many
epoch of reionization experiments are in fact intensity mapping experiments,
and the same method can also be used to observe large scale structure at
the redshift range $0<z<3$, which is predicted to be more sensitive to cosmological
parameters than current galaxy surveys \citep{2008PhRvL.100i1303C,2009astro2010S.234P}.
This method is tested with existing
telescopes such as the Green Bank Telescope (GBT) and the Parkes telescope, 
and has detected positive
cross-correlation between the 21 cm and optical
galaxies \citep{2010Natur.466..463C,2013ApJ...763L..20M,2013MNRAS.434L..46S}.
However, the time available on these general purpose telescopes are limited.
It has been argued that large cylinderical reflectors can be cheaply made
and used in hydrogen survey experiments \citep{2006astro.ph..6104P}.

In a cylinder array designed for this purpose, a number of cylinder reflectors
are fixed on the ground pointing the zenith,
along the north-south direction and in parallel with each other.
Receiver feeds are placed along the focus line of each of the cylinders,
forming an interferometer array. As the Earth rotates,
the array will drift scan the visible sky.
For the observation of the BAO features
in the large scale structure, which can be used to probe dark energy
properties, a cylinder array of 100m-150m size would be
optimal \citep{2008arXiv0807.3614A,2012A&A...540A.129A,2010ApJ...721..164S}.

The Tianlai project \footnote{\url{http://tianlai.bao.ac.cn}. The word
{\it Tianlai} means ``heavenly sound'' in Chinese, this phrase appeared
first in the work of ancient Chinese philosopher Chuang Tzu (369BC-286BC).}
is an experimental effort in this direction \citep{2012IJMPS..12..256C,2011SSPMA..41.1358C}.
Tentatively, we shall assume in this paper that the full-scale Tianlai experiment
will consist of 8 adjacent cylinders, each with 15m wide and 120m long, totally with about 2000 dual
polarization units, covering the frequency range of  $400-1420\MHz$, corresponding
to $0<z<2.5$. We shall assume a system temperature of $50 \K$.

At present, a pathfinder experiment is being built in a radio quiet site
at Hongliuxia, Balikun County, Xinjiang Autonomous Region,
China. 
This pathfinder consists of both a cylinder array and a dish array. The dish array
will include 16 steerable 6 m dishes, and it will be discussed in a separate paper, here we
shall focus on the cylinder array.
The cylinder pathfinder array include three adjacent cylindrical reflectors, each with
15 m wide and 40 m long. It will focus on  observing at the frequency range of
700 -- 800 MHz. Currently, the pathfinder cylinders have a total of 96 receivers, averaging
32 on each. 
With a margin of 5 meters on each of the two ends of the cylinder, the distance between
the feeds is around 97 cm, greater than one wavelength at $z=1$ ($\lambda_{\rm obs}=21(1+z) \cm$).
After a period of experiment, we plan to expand the total number of dual polarization
 receivers later to 216, so that on each cylinder there are on average 72 dual polarization
receivers. We shall call this the {\it pathfinder+} experiment.
Using the pathfinder and pathfinder+ experiment, we hope to demonstrate the feasibility
of intensity mapping with the cylinder array, before
building the full-scale experiment. These configuration parameters for the
cylinder pathfinder, pathfinder+ and full-scale experiment
are listed in Table \ref{tab:exp}.

Another potentially interesting application of a 21 cm intensity mapping experiment is to
look for and constrain the primordial non-Gaussianity.  
The primordial density perturbations, which were originated during the inflation era 
and gave rise to the various structures today, link the observable Universe to the very 
early phase of the Universe.
While the simplest slow-roll inflation model predicts very weak primordial non-Gaussianity in 
the density perturbations with an amplitude below the detectable level, many other inflation 
mechanisms could result in observable non-Gaussianities 
(see \citealt{2004PhR...402..103B,2010AdAst2010E..72C} for reviews).
Any observational constraint on the level of the primordial non-Gaussianity 
can be a powerful probe to the dynamics of inflation.

Various observational approaches, such as the angular bispectrum of the CMB, 
high-order correlations of the three dimensional 
galaxy distribution, abundance of rare objects, and the 
large-scale clustering of halos, have been developed to constrain the level of 
primordial non-Gaussianity, specifically the nonlinearity parameter $f_{\rm NL}$ 
(see \citealt{2010AdAst2010E..73L,2010AdAst2010E..64V} for reviews).
Combining large scale clustering measurements from galaxy surveys with their 
cross-correlations with the CMB from the Wilkinson Microwave Anisotropy Probe (WMAP) 
nine-year data, \citet{2014PhRvD..89b3511G} obtained $-37 < f_{\rm NL} < 25 $ at $95\%$ 
confidence for the local-type configuration.
The latest and tightest constraints on $f_{\rm NL}$ come from the measurement of the 
CMB angular bispectrum by Planck, which are $f_{\rm NL}^{\rm local} = 2.7 \pm 5.8$, 
$f_{\rm NL}^{\rm equil} = -42 \pm 75$, and $f_{\rm NL}^{\rm ortho} = -25 \pm 39$ (68\% CL) 
for the primordial local, equilateral, and orthogonal bispectrum amplitudes 
respectively \citep{2013arXiv1303.5084P}.
Using the large scale clustering of tracers of dark matter in the later Universe, two
most commonly used probes for the primordial non-Gaussianity are the scale-dependent bias
in the observed power spectrum, and the bispectrum.

In this paper we make simple forecasts on the constraining power of the Tianlai experiment,
under the assumption of perfect foreground removal and no systematics.
We shall make our forecasts primarily for the full-scale experiment, which is
designed to measure the large scale structure and cosmological parameters. We will
also make some forecasts on the pathfinder and pathfinder+ experiments, which are
only used for testing the key technology for the full-scale experiment and are
not expected to achieve any good precision.

The paper is organized as follows: in Sec.2, we present  
the signal power spectrum, as well as the detailed formalism for estimating
the noise power spectrum for an interferometer array, and forecast the measurement
error of the power spectrum by the Tianlai arrays. 
Based on the power spectrum measurement, we
forecast the constraints on dark energy parameters obtainable from Tianlai
BAO and RSD observations in Sec.3. In Sec.4, we briefly review the imprint of
primordial non-Gaussianity on the large scale structures,
in Sec.4.1 we study the constraint obtainable by
considering the scale-dependent bias in the power spectrum, and in Sec.4.2 we apply
the bispectrum method. We conclude in Sec.5.

\begin{center}
\begin{table}[htp]
\caption{\label{tab:exp} The experiment parameters for Tianlai.}
\begin{tabular}{cccccc}
\hline
 & cylinders & width & length & dual pol. units/cylinder & Frequency \\
\hline
Pathfinder & 3 & 15 m & 40 m & 32 & 700 -- 800 MHz \\
\hline
Pathfinder+ & 3 & 15 m & 40 m & 72 & 700 -- 800 MHz \\
\hline
Full scale & 8 & 15 m & 120 m & 256 & 400 -- 1420 MHz \\
\hline
\end{tabular}
\end{table}
\end{center}

\section{The Power Spectrum Measurement with Tianlai Intensity Mapping}



\subsection{The Signal Power Spectrum}\label{P_signal}

In an HI intensity mapping observation, the distances along and perpendicular to
the line of sight are measured from redshift and angular separation respectively,
and the HI power spectrum is observed in redshift space. Therefore, the observed HI 
power spectrum is given by \citep{2003ApJ...598..720S}
\begin{eqnarray}
\label{eqa:Pobs}
P_{\rm obs}(k_{\rm ref\perp},k_{\rm ref\parallel})&=&\frac
{D_A(z)_{\rm ref}^{2}H(z)}{D_A(z)^{2} H_{\rm ref}(z)}
\left(b_1^{\rm HI}(z)+f(z)\,\frac{k_{\parallel}^{2}}{k_{\perp}^{2}
+k_{\parallel}^{2}}\right)^{2} 
\times G(z)^{2}P_{\rm m0}(k)+P_{\rm shot},
\end{eqnarray}
where the subscript ``ref'' denotes quantities calculated in the
reference cosmology, for which we use the Planck 2013 parameters \citep{2013arXiv1303.5076P}.
Here $b_1^{\rm HI}(z)$ is the linear bias factor of the HI gas at redshift $z$, 
$f(z)$ is the linear growth rate,
$k_{\perp}$ and $k_{\parallel}$ are the $\bf k$
components perpendicular and parallel to the line of sight, respectively, 
$G(z)$ is the growth factor, $P_{\rm m0}(k)$ is the present matter power spectrum, and
$P_{\rm shot}$ is the shot noise contribution.

In a model with dark energy equation of state
$w(z)$, the hubble parameter and the angular diameter distance are given by
\begin{equation}
\label{eqa:HubZ}
\frac{H(z)}{H_0}=\left[\Omega_m(1+z)^3+  \Omega_k(1+z)^2 +
\Omega_X e^{3\int_0^z\frac{1+w(z')}{1+z'}dz'} \right]^{1/2}
\end{equation}
and
\begin{equation}
\label{eqa:DaZ}
D_A(z)=\frac{c}{1+z}\int_0^z\frac{dz'}{H(z')} .
\end{equation}
Thus the present density and the equation of state parameters of dark energy can be constrained
by measuring the acoustic peaks on the power spectrum.

The redshift space distortion of the power spectrum also provides information
on the growth history of the Universe.
The linear growth rate $f(z)$ affects the
observed power spectrum (Eq.~(\ref{eqa:Pobs})) through the
redshift space distortion (RSD) factor $\beta$, by 
\begin{equation}
\beta \,=\, f(z)/b_1^{\rm HI}(z), 
\end{equation}
and through the linear growth factor $G(z)$, 
which is related to $f(z)$ by 
\begin{equation}
f =\frac{d\ln G(a)}{d\ln a} = -\frac{(1+z)}{G(z)}\frac{d G(z)}{dz}.
\end{equation}
As the growth factor $G(z)$ is degenerate with the HI bias factor, here we focus on the
growth rate obtained from the redshift space distortion, and will discuss 
the measurement error on $f(z)$.

The redshift space power spectrum measured from the 21 cm intensity mapping
could also be used as a test for gravity \citep{2013PhRvD..87f4026H,2010PhRvD..81f2001M},
or provide extra information on the dark energy if general relativity is assumed.
For dark energy models, the growth rate can be parameterized as $f(z) = \Omega_{\rm m}^{\gamma}(z)$,
with $\gamma_{\rm \Lambda CDM} \approx 0.55$ for the $\Lambda$CDM+GR model. 
The value of $\gamma$ in other dark energy
models with $w$ different from $-1$ does not deviate from $\gamma_{\rm \Lambda CDM}$ significantly.

The intensity mapping observation directly measures the 21 cm brightness temperature, and
the measured 21 cm power spectrum, $P_{\Delta T}(\vec{k}) = \bar{T}_{\rm sig}^2 P_{\rm obs}(\vec{k})$, 
is the power spectrum of brightness temperature $\delta T_b$ due to the 21 cm emission,
in which the average signal temperature $\bar{T}_{\rm sig}$ has been
estimated \citep{2007RPPh...70..627B,2008PhRvL.100i1303C,2010ApJ...721..164S} to be
\begin{equation}
\bar{T}_{\rm sig} \;=\; 190\, \frac{x_{\rm HI}(z)\,\Omega_{\rm H,0}\,h\,(1+z)^2}{H(z)\,/\,H_0} \mK,
\end{equation}
where $x_{\rm HI}(z)$ is the neutral fraction of hydrogen at
redshift $z$, and $\Omega_{\rm H,0}$ is the ratio of the hydrogen mass
density to the critical density at $z = 0$. 


After the completion of cosmic reionization, the HI gas in the Universe is mostly distributed in galaxies hosted by halos. Therefore, we model the HI bias factors as halo bias factors weighted by the neutral hydrogen mass hosted by these halos \citep{2011ApJ...740L..20G}:
\begin{equation}\label{bHI}
b_i^{\rm HI}(z) \;=\; \frac{\int_{M_{\rm min}}^{M_{\rm max}}\, dM\, n(M,z)\, M_{\rm HI}(M)\, b_i(M,z)}{\rho_{\rm HI}},
\end{equation}
for $i=1$ and $2$, where $\rho_{\rm HI}$ is the mass density of HI gas, $n(M,z)$ is the halo mass function for which we use Sheth \& Tormen's formalism \citep{1999MNRAS.308..119S}, $M_{\rm HI}(M)$ is the HI mass in a halo of mass $M$, and $b_1(M,z)$ and $b_2(M,z)$ are halo bias parameters. The mass density of HI clouds is given by
\begin{equation}\label{Eq.rho_HI}
\rho_{\rm HI} \;=\; \int_{M_{\rm min}}^{M_{\rm max}}\, dM\, n(M,z)\, M_{\rm HI}(M).
\end{equation}
Following \citet{2011ApJ...740L..20G}, we take $M_{\rm min} = 10^8 \h^{\rm -1} \Msun$ for halos to retain their neutral gas \citep{2001ARA&A..39...19L}, and take $M_{\rm max} = 10^{13} \h^{\rm -1} \Msun$ for the gas to have sufficient time to cool and form galaxies.


As for the relation between the HI gas mass $M_{\rm HI}$ and the host halo mass $M$, we use the fitting result by \citet{2011ApJ...740L..20G}, which is based on numerical simulation and consistent with observations:
\begin{equation}
M_{\rm HI}(M) \;=\; A \times \left(1\,+\, \frac{M}{c_1}\right)^b\, \left(1 \,+\, \frac{M}{c_2}\right)^d,
\end{equation}
for $M > 10^{10} \Msun$, and $M_{\rm HI}(M) = X_{\rm HI}^{\rm gal}\,(\Omega_{\rm b}/\Omega_{\rm m})\, M$ with $X_{\rm HI}^{\rm gal} = 0.15$ for $M \le 10^{10} \Msun$.
The best-fit parameters are $A = 2.1\times 10^8$, $c_1=1.0\times10^{11}$, $c_2=4.55 \times 10^{11}$, $b=2.65$, and $d=-2.64$ for redshift $z=1$.
As the $M_{\rm HI}$ - $M$ relation does not change much from $z=1$ to $z=3$ \citep{2011ApJ...740L..20G}, we use the fixed values of these parameters throughout our calculation.

The halo bias factors can be obtained from the halo model (see \citealt{2002PhR...372....1C} for a review).
The linear and the first non-linear bias factors of halos are \citep{2001ApJ...546...20S,1997MNRAS.284..189M}
\begin{equation}
b_1(M,z) \;=\; 1 \,+\, \epsilon_1 \,+\, E_1,
\end{equation}
\begin{equation}
b_2(M,z) \;=\; 2\,(1 \,+\, a_2)\, (\epsilon_1 \,+\, E_1) \,+\, \epsilon_2 \,+\, E_2,
\end{equation}
where
\begin{equation}
\epsilon_1 \;=\; \frac{q\nu - 1}{\delta_{\rm sc}(z)}, \;\; \epsilon_2 \;=\; \frac{q\nu}{\delta_{\rm sc}(z)} \, \left(\frac{q\nu - 3}{\delta_{\rm sc}(z)}\right),
\end{equation}
and
\begin{equation}
E_1 \;=\; \frac{2p/\delta_{\rm sc}(z)}{1 + (q\nu)^p}, \;\; \frac{E_2}{E_1} \;=\; \frac{1 + 2p}{\delta_{\rm sc}(z)} \,+\, 2\epsilon_1.
\end{equation}
Here $a_2 = -17/21$, $\nu \equiv \delta_{\rm sc}^2(z)/\sigma^2(M)$, and $\delta_{\rm sc}(z)=1.686/G(z)$ is the critical overdensity required for spherical collapse at $z$, extrapolated to the present time using linear theory. For Sheth \& Tormen's halo mass function \citep{1999MNRAS.308..119S}, $p\approx 0.3$, and $q = 0.707$.



\subsection{Generalized  Noise Power Spectrum}\label{gen_formalism}
The fundamental observable of a radio interferometer is the visibility,
which is the correlation between outputs of two receivers for a given baseline.
For a given sky brightness distribution $I(\hat{n}, \nu)$,
where $\hat{n}$ and $ \nu$ are the
sky position and the observing frequency respectively, the
corresponding visilibity, in units of flux density, can be written as
the Fourier transform of the sky brightness
weighted by the beam pattern $A(\hat{n})$ of the two receivers:
\begin{eqnarray}
\label{eq:visibi_Jy}
V_{\alpha\beta,\rm [Jy]} (\vec{u}_{\alpha\beta}, \nu)
&=& \int d^2\hat{n} ~ e^{-i 2\pi \hat{n} \cdot \vec{u}_{\alpha\beta} } A_{\alpha}(\hat{n},\nu)
A^{\ast}_{\beta}(\hat{n},\nu) I(\hat{n}, \nu) \nonumber \\
&\approx& \int d^2\hat{n} ~ e^{-i 2\pi \hat{n} \cdot \vec{u}_{\perp} } A_{\alpha}(\hat{n},\nu)
A^{\ast}_{\beta}(\hat{n},\nu) I(\hat{n}, \nu),
\end{eqnarray}
where $\vec{u}_{\alpha\beta}$ denotes the baseline vector in units of wavelength.
Here in the second equality, we have used the flat-sky approximation, and
$\vec{u}_{\perp}$ is the component of $\vec{u}_{\alpha\beta}$ perpendicular
to the line of sight.
In a large-scale survey like Tianlai, the flat-sky assumption will
certainly break down, a full-sky representation based on the
spherical harmonic expansion has been developed \citep{2014ApJ...781...57S}.
Here we use the flat-sky approximation and Fourier expansion,
as for forecasting it is still sufficient.

For radio interferometers, it is convenient to define the equivalent
visibility in units of brightness temperature, using the Rayleigh-Jeans approximation,
so that
\begin{equation}
\label{eq:visibi_K}
V_{\alpha\beta,\rm [K]} (\vec{u}_{\perp}, \nu)
\;=\; \int d^2\hat{n} ~ e^{-i 2\pi \hat{n} \cdot \vec{u}_{\perp} } A_{\alpha}(\hat{n},\nu)
A^{\ast}_{\beta}(\hat{n},\nu) \delta T_b(\hat{n}, \nu).
\end{equation}
The thermal noise of the measurement can be written as 
\begin{eqnarray}
\delta V_{\alpha\beta,\rm [K]} (\vec{u}_{\perp}, \nu)
= \frac{\lambda^2\, T_{\rm sys}}{A_e\sqrt{\Delta \nu t_{\vec{u}}}},
\end{eqnarray}
where $\Delta \nu$ is the observed full bandwidth,
$t_{\vec{u}} $ is the integration time of this baseline, 
$T_{\rm sys}$ is the system temperature per polarization 
(we assume $T_{\rm sys}=50 \K$ in this paper),
and $A_e$ is the effective collecting area of each element. We can make a further Fourier
transform of the visibility with respect to $\nu$, to obtain the so called 
visibility delay-spectrum\citep{2012ApJ...756..165P},
\begin{equation}
\label{eq:visibi_KMHz}
V_{\alpha\beta,\rm [K\cdot MHz]} (\vec{u}_{\perp}, u_{\parallel})
\;=\; \int d\nu~ e^{-i 2\pi \nu u_{\parallel}} V_{\alpha\beta,\rm [K]} (\vec{u}_{\perp}, \nu).
\end{equation}
Now the three-dimensional vector $\vec{u} \equiv \{ \vec{u}_{\perp}, u_{\parallel} \}$
is the Fourier conjugate of the sky position vector
$\vec{\theta}=\{ \hat{n}, \nu \}$.
The thermal noise in this representation  is then \citep{2005ApJ...619..678M}
\begin{eqnarray}
\Delta T_N (\vec{u})
\;=\; \frac{T_{\rm sys}}{\sqrt{\Delta\nu t_{\vec{u}}}}\left( \frac{\lambda^2 \Delta\nu}{A_e}\right).
\end{eqnarray}
Here the factor $\lambda^2 \Delta\nu/A_e$ represents the Fourier 
space resolution of the observation
in the sense that any two vectors within it will be highly correlated.

To extract cosmological information, one is interested in the correlation 
function of the visibilities measured
at discrete baselines $\vec{u}_i$ and $\vec{u}_j$. 
If we neglect the correlation of thermal noise errors between measurements, 
the noise covariance matrix for visibilities is approximately diagonal, and can
be written as \citep{2006ApJ...653..815M,2003JApA...24...23B}
\begin{eqnarray}
\label{eqn:noise_covm}
\bfC^{N}(\vec{u}_i, \vec{u}_j) = \langle \Delta T_N(\vec{u}_i) \Delta T^{\ast}_N(\vec{u}_j)\rangle
= \left ( \frac{\lambda^2 \,T_{\rm sys} \Delta \nu }{A_e}
\right)^2   \frac{\delta_{ij}}{\Delta \nu \, t_{\vec{u}}}.
\end{eqnarray}
The integration time for baseline $\vec{u}$ can be written
as 
\begin{equation}
t_{\vec{u}} = \frac{A_e}{\lambda^2}
 n(\vec{u}_{\perp}) t_{\rm int},
\end{equation}
where $n(\vec{u}_{\perp})$ is the baseline number density of
the interferometer in $u-v$ plane, and
$A_e/\lambda^2 \approx \delta u\, \delta v$ is the $\vec{u}$-space resolution.
For an observation with survey area $\Omega_{\rm map}$ larger
than the field of view $\Omega_{\rm FOV}$ and uniform survey coverage,
the integration time of each pointing
$t_{\rm int}= t_{\rm tot} (\Omega_{\rm FOV}/\Omega_{\rm map}) $.

The the sample variance contribution to the covariance matrix is
\citep{2006ApJ...653..815M}
\begin{eqnarray}
\bfC^{SV}(\vec{u}_i, \vec{u}_j) = \langle \delta T_{b}(\vec{u}_i) \delta  T^{\ast}_{b}
(\vec{u}_j ) \rangle
&\approx &   \delta_{ij} \int d^3\vec{u} ~ |R(\vec{u}_i -\vec{u})|^2 ~ P_{\Delta T}(\vec{u})
\nonumber \\
&\approx & \delta_{ij}  \frac{\lambda^2 \Delta \nu^2}{r_{a}^2(z) \Delta r(z) A_e}
P_{\Delta T}(\vec{k}_{i \perp}, k_{i \parallel}),
\end{eqnarray}
where $P_{\Delta T}$ is the 21 cm signal power spectrum.
Here $R(\vec{u}_i - \vec{u})$ is the response function for a given baseline $\vec{u}_i$,
which is defined as the Fourier transform of the primary beam
$A_{\alpha}(\hat{n},\nu) A^{\ast}_{\beta} (\hat{n},\nu)$ in Eq.\ (\ref{eq:visibi_Jy}).
The Kronecker $\delta_{ij}$ arises due to the choice of the pixel
size that is approximately the same as
the support of function $R(\vec{u})$. The integration of $|R|^2$
then introduces a factor approximately
equals the inverse of the Fourier space resolution, $\lambda^2\Delta\nu/A_e$,
due to the normalization of $R(\vec{u})$.
Here $\Delta r = y(z) \Delta \nu$ is the spatial resolution
corresponding to bandwidth $\Delta \nu$.
The comoving angular diameter distance $r_a(z)$ and
the factor $y(z)= \lambda_{21} (1+z)^2/ H(z)$ are used to
convert the power spectrum from $\vec{u}$-space to the comoving $\vec{k}$-space:
\begin{eqnarray}
\vec{u}_{\perp} = \frac{r_a(z) k_{\perp}}{2\pi}, \qquad
u_{\parallel}  = \frac{y(z) k_{\parallel}}{2\pi}.
\end{eqnarray}

Given the total covariance matrix $\bfC=\bfC^N+\bfC^{SV}$,
one could then estimate the measurement uncertainty
of the bandpower from the Fisher matrix
\begin{eqnarray}
F_{ab} = {\rm Tr}\left [ \bfC^{-1}\frac{\partial \bfC}{\partial p_a}
\bfC^{-1}\frac{\partial \bfC}{\partial p_b} \right],
\end{eqnarray}
where parameter $p_a$ is the bandpower $p_a = P_{\Delta T} (\vec{k}_a)$.
For diagonal $\bfC$, the measurement error $\delta P_{\Delta T}$ is
\begin{eqnarray}
\delta P_{\Delta T}  (\vec{k}_i)  = \frac{1}{\sqrt{N_c}(\vec{k}_i)}
\frac{A_e r_a^2 \Delta r}{\lambda^2 \Delta \nu^2} \left [C^N(\vec{k}_i, \vec{k}_i) +
C^{SV}(\vec{k}_i, \vec{k}_i) \right]
= \frac{1}{\sqrt{N_c}(\vec{k}_i)} \left[P^{N}(\vec{k}_i) + P^{SV}(\vec{k}_i) \right].
\end{eqnarray}
where the number of modes 
$N_c(k) \,=\, k_{\perp} dk_{\perp} dk_{\parallel} \, V/(2\pi)^2$,
with $V$ being the survey volume. Here we have denoted the signal power spectrum
in the sample variance term as the sample variance power spectrum, i.e. 
$P^{SV}(\vec{k}_i) = P_{\Delta T}(\vec{k}_i)$, and
the noise power spectrum $P^N(\vec{k})$ is
\begin{eqnarray}
P^{N}(k, z) = \frac{4\pi f_{\rm sky} \lambda^2\, T^2_{\rm sys}\, y(z)\, r_a(z)^2}
{A_e\, \Omega_{\rm FOV}\, t_{\rm tot} } \left( \frac{\lambda^2}{A_e\, n(\vec{k}_{\perp}) }\right),
\end{eqnarray}
where $f_{\rm sky}$ is the fraction of the sky coverage, i.e. $f_{\rm sky}=\Omega_{\rm map}/4\pi$,
and $\Omega_{\rm FOV}$ is the field-of-view of a single pointing.

\subsection{Tianlai Noise Power Spectra}

We first calculate the baseline distribution function $n(\vec{u}_{\perp})$ of
the interferometer. In a real interferometer, for a pair of antennae with
separation $\vec{u}$, the output is actually the average of the visibility
on a region of the u-v plane centered at $\vec{u}$.
Instead of the discrete histogram, therefore, we incorporate the
response function of an antenna pair $R(\vec{u})$ \citep{2012A&A...540A.129A}
and derive a continuous function $n(\vec{u}_{\perp})$, with the caveat that
only $n(\vec{u}_{\perp}) (A_e/\lambda^2)$ is physically meaningful in this formalism. 
For the Tianlai cylinder array with receivers fixed along the focal lines of cylinders,
the pair response pattern of a cylinder can be approximated as 
a two-dimensional triangular function with rectangular support
\citep{2001isra.book.....T,2008arXiv0807.3614A,2012A&A...540A.129A},
which is set by the cylinder width $W$ in east-west
direction, $\Delta u_W = W/\lambda$, and the feed length $L$
in north-south direction, $\Delta u_L= L/\lambda$:
\begin{eqnarray}
R(\vec{u}_{\perp}) = \left ( \frac{\lambda^2}{A_e} \right)
\Lambda\left(\frac{u_L}{\Delta u_L}\right) \Lambda \left( \frac{u_W}{\Delta u_W} \right) .
\end{eqnarray}
Here the triangular function $\Lambda(x)$ is defined as
$1-|x|$ for $|x|<1$ and $0$ otherwise.
The baseline number density $n(\vec{u}_{\perp})$ could be obtained simply by summing up $R(\vec{u})$ 
for all baselines, i.e.
\begin{eqnarray}
\label{eqn:nu}
n(\vec{u}_{\perp}) = \sum_{i}^{n_b} R(\vec{u}_{\perp} - \vec{u}_{\perp}^i ).
\end{eqnarray}
The baseline number density $n(\vec{u}_{\perp})$ is normalized that the half-plane integral
would give the total baseline number of $n_b=n_r(n_r-1)/2$,
where $n_r$ is the total number of receivers.

\begin{figure}[ht]
\centering{
\includegraphics[scale=0.42]{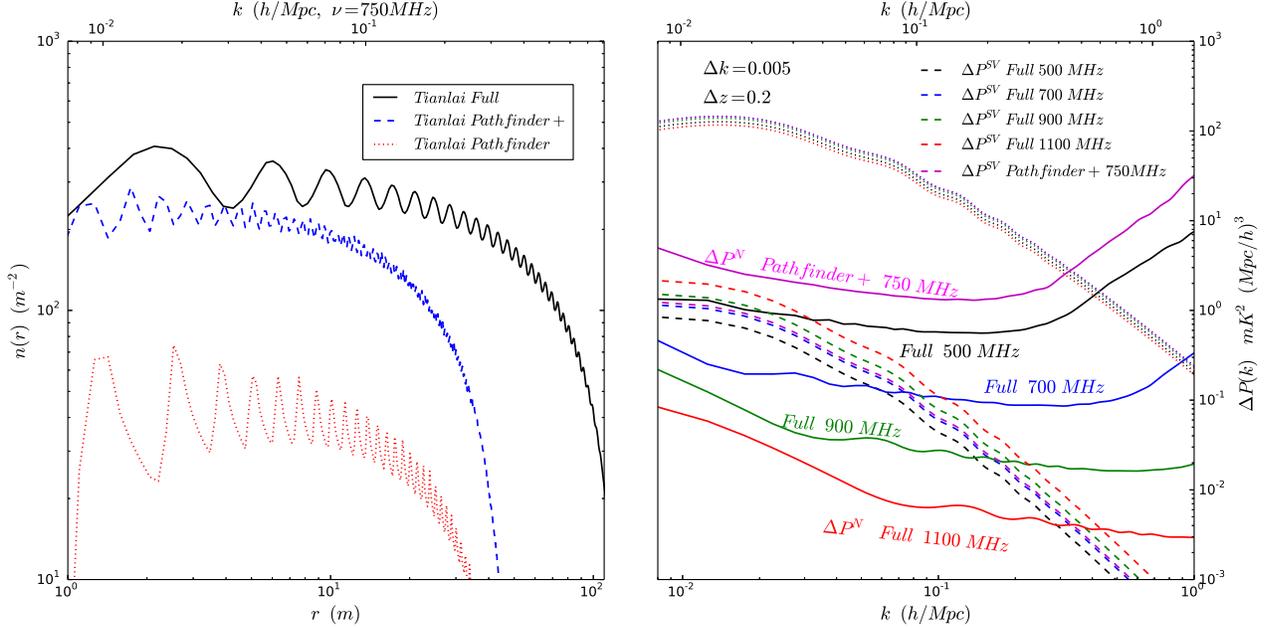}
\caption{ ({\it Left}): one-dimensional baseline distribution $n(r)$ of various Tianlai configurations, 
calculated from Eq.\ (\ref{eqn:nu}). Here x-axis $r$ is the physical distance in the interferometer frame, 
we also display the corresponding cosmological scale it probes at frequency $\nu=750 \MHz$ on the upper abscissa.
({\it Right}): The measurement error contributed from thermal noise, $\Delta P^N(k) = P^N(k)/\sqrt{N_c(k)}$ 
({\it solid lines}), and that from sample variance, $\Delta P^{SV}(k) = P^{SV}(k)/\sqrt{N_c(k)}$ ({\it dashed lines}).
The case for Tianlai pathfinder+ is shown as the magenta lines, while the cases for  
full-scale Tianlai at various frequencies are shown as other colors.
The 21 cm signal power spectra, $P_{\Delta T}(k)$, 
at corresponding redshifts are shown by dotted lines with corresponding colors,
assuming a constant HI fraction $x_{\rm HI}=0.008$ at all redshifts.
Here we adopt a wavenumber bin width of $\Delta k = 0.005\Mpc/\h$, and $\Delta z = 0.2$.}
\label{Fig.pnoise}
}
\end{figure}

In the left panel of Fig.~\ref{Fig.pnoise}, we plot the baseline
distribution $n(r)$ of different configurations of Tianlai. Besides
the baseline distance $r$ in the interferometer frame, we also show
on the upper abscissa the cosmological scales that the array could
probe at frequency $\nu=750 \MHz$.
Given the same cylinder dimension of pathfinder and its upgrade,
the baseline densities of these two configurations drop
at similar scales, around $20\m$. However, the larger feeds number
of the pathfinder+ would increase the number
of baselines available at a given scale.  For the full-scale Tianlai,
both the numbers of shorter and longer baselines
are increased, therefore, the sensitivity will be improved significantly.

In the right panel of  Fig.~\ref{Fig.pnoise}, we plot the measurement error on the 
power spectrum due to thermal noise $\Delta P^N(k) = P^N(k)/\sqrt{N_c(k)}$
({\it solid lines}) and sample variance 
$\Delta P^{SV}(k) = P^{SV}(k)/\sqrt{N_c(k)}$ ({\it dashed lines}).
While only the errors at the medium frequency $f=750 \MHz$ are shown (the magenta set of lines)
for the pathfinder+, we display four different frequencies from $500 \MHz$
to $1100 \MHz$ (from top to bottom for the thermal noise, and from bottom to top
for the sample variance) for the full-scale survey.
For the sample variance power spectra, 
we adopt a conservative assumption that $x_{\rm HI}=0.008$ at all redshifts.
For comparison, we also plot the 21 cm signal power spectra, $P_{\Delta T}(k)$, 
at the corresponding redshifts with the dotted lines.

From the figure, we can see that for Tianlai pathfinder+, the  thermal noise will 
dominate over the sample variance at all scales.
This is also true for the high-redshift observation of the full-scale Tianlai, but
the thermal noise gradually decreases towards lower redshift. At $z\sim 1$ (Full 700MHz),
the two contributions are comparable at the BAO scale.

\subsection{The Power Spectrum with Expected Tianlai Errors}

As discussed in Sec.\ref{gen_formalism}, the measurement error of the power spectrum 
is a sum of the sampling error and thermal noise.
Since the measured 21 cm power spectrum is proportional to the HI power spectrum by a factor
of $\bar{T}_{\rm sig}^2$, we use the measurement error on the HI power spectrum
for the Fisher forecasts in the following sections, and it is 
\begin{equation}
\Delta P_{\rm obs} (\vec k) \;=\; \frac{1}{\sqrt{N_c}}\, \left[ P_{\rm obs}(\vec k) + N(k)\right],
\end{equation}
where $N(k)$ is related to the thermal noise power by 
$P^N = \bar{T}_{\rm sig}^2\, N(k)$, 
and $N_c$ is the number of independent modes in that pixel in Fourier space
as discussed in section \ref{gen_formalism}. 

\begin{figure}[ht]
\centering{
\includegraphics[scale=0.35]{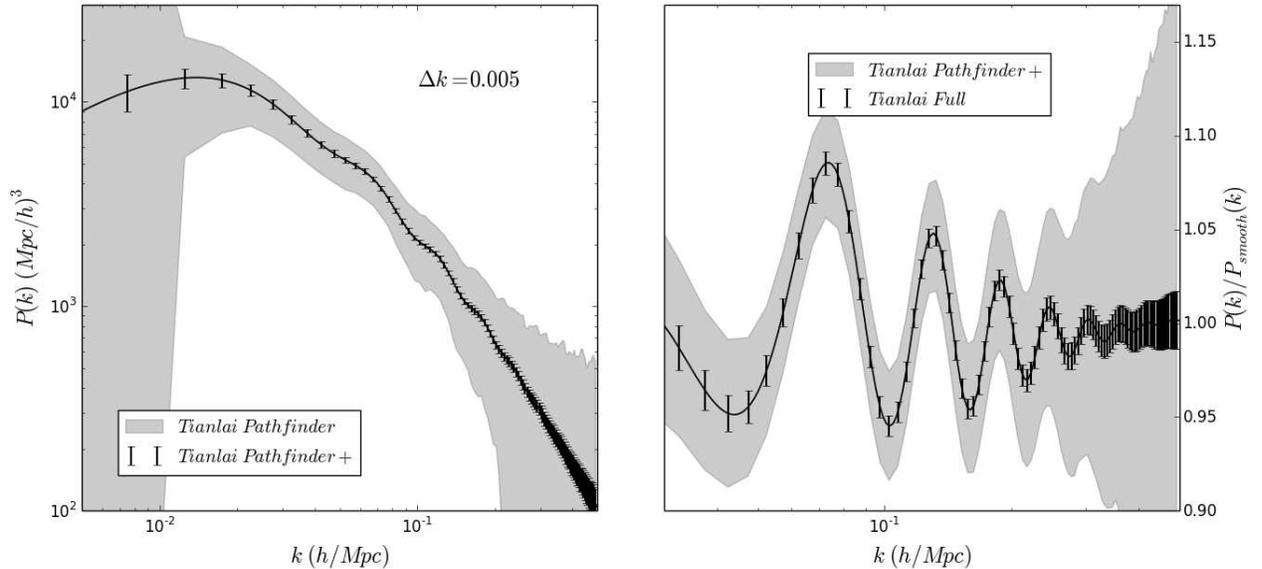}
\caption{{\it Left panel:} the measurement errors on the power spectrum at $z=1$ for the
Tianlai pathfinder ({\it shaded area}) and the pathfinder+ ({\it error bars}).
{\it Right panel:} the relative one with respect to the smooth
power spectrum with errors expected from the Tianlai pathfinder+ ({\it shaded area}) and
the full-scale Tianlai ({\it error bars}). The assumed survey area is 10000 $deg^2$, and 
the integration time is 1 year. The wavenumber bin width for this plot is 
$\Delta k = 0.005\Mpc/\h$.}
\label{Fig.pk_error}
}
\end{figure}

The instantaneous field of view of a cylindrical radio telescope is narrow in right ascension
but very broad in declination, limitted primarily by the illumination angle of the feeds.
The rotation of the Earth results in a broad coverage in right ascension.
The illumination angle of the feeds of Tianlai is about $120^\circ$.
Assuming a latitude of $\theta_{\rm lat} = 44^\circ$, the Tianlai array covers the
declination angle from $-16^\circ$ to $90^\circ$.
Considering the masking effects in order to avoid the disk area of the Milky Way and
bright radio sources, we conservatively assume a survey area of 10000 $\deg^2$ throughout.

The left panel of Fig.~\ref{Fig.pk_error} shows the power spectrum at $z=1$ with measurement errors
expected from the Tianlai pathfinder ({\it shaded area}) and the 
pathfinder+ ({\it error bars}), while
the right panel shows the relative power spectrum with respect to the smooth power spectrum with
errors expected from the Tianlai pathfinder+ ({\it shaded area}) and the full-scale Tianlai ({\it error bars}) respectively. The integration time is assumed to be 1 year. Note that the error bar for 
$P(k)$ depends on the binning of $k$. Here we have chosen a bin width of $\Delta k = 0.005\Mpc/\h$.
If a different bin width is chosen, the error bar on power spectrum can be obtained by scaling, 
but the constraints on our interested parameters are insensitive to the choice of binning.

\section{Fisher forecast on the constraint on dark energy}\label{fisher_DE}

From the power spectrum measurement at a given redshift, the 
Fisher information matrix can be written 
as \citep{1997PhRvL..79.3806T,2003ApJ...598..720S,2008PhRvD..78b3529M}
\begin{equation}\label{Eq.fisher_PS}
F_{\alpha \beta} \;=\; \sum_k\, \left[ \frac{\partial P_{\rm obs}(\vec k)}{\partial \alpha}\,
\frac{\partial P_{\rm obs}(\vec k)}{\partial \beta}\right]\,/\, \left[ \Delta P_{\rm obs} (\vec k) \right]^2.
\end{equation}
Here the free parameters $\alpha$ and $\beta$ are taken from $\{D_{\rm A}(z_i), H(z_i), b_{1,i}^{\rm HI}, 
f(z_i), P_{\rm shot,i}\}$ for each redshift bin $z_i$.
The nuisance parameters in the model $\{b_{1,i}^{\rm HI}, P_{\rm shot,i}\}$ can be marginalized 
by selecting the submatrix of $F^{-1}_{\alpha \beta}$ with only the appropriate columns and rows.
We can then derive the measurement errors on the expansion and structure growth history parameters.

For the Tianlai pathfinder and pathfinder+ experiment, the observation frequency range is 
700 -- 800 MHz, and we divide this frequency band into 3 bins with equal bandwidth, and obtain 
estimates of measurement error for the corresponding redshift bins. For 
the planned full-scale Tianlai experiment, the 
frequency range of 400 -- 1420 MHz is equally divided into 8 bins, the bin size in 
this latter case is larger than in the pathfinder case, but 
the bin size $\Delta z$ does not really matter in the end.

\begin{figure}[t]
\centering{
\includegraphics[scale=0.38]{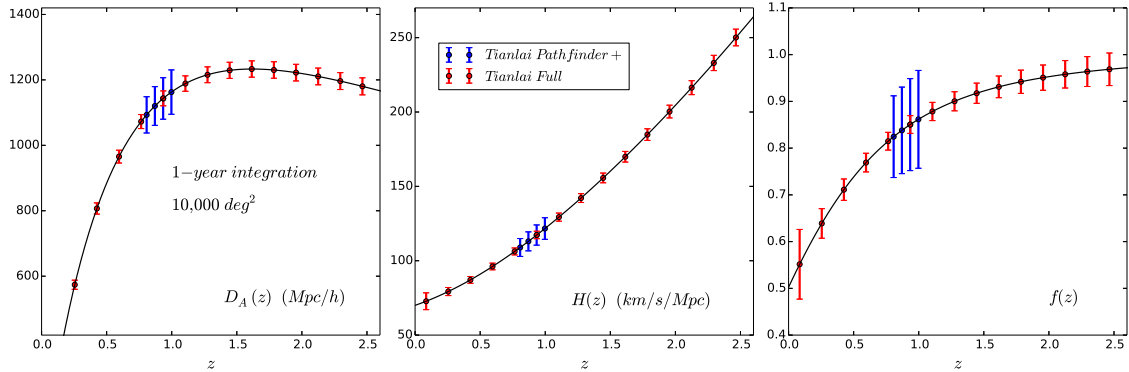}
\caption{Measurement errors on the angular diameter distance $D_{\rm A}$ ({\it left panel}),
the Hubble expansion rate $H$ ({\it central panel}), and the growth rate
$f(z)=d\ln G / d \ln a$ ({\it right panel}). The integration time is assumed to be 1 year.
Note that here the number of redshift bins is larger than
the one we used for forecasting the constraint on dark energy equation of state,
just to make the error bars visible.}
\label{Fig.DA_H_error}
}
\end{figure}

We plot the measurement errors on the angular diameter distance $D_{\rm A}(z_i)$, the Hubble
expansion rate $H(z_i)$, and the growth rate
$f(z_i)$ in the left, central, and right panels of Fig.~\ref{Fig.DA_H_error}, respectively,
for the pathfinder+ (blue error bars) and the full-scale Tianlai experiments
(red error bars). The integration time is assumed to be 1 year in all cases.
We see that the pathfinder+ experiment can obtain a useful measurement on 
$D_A(z)$ and $H(z)$ at $z=1$. For the growth rate $f$, the errors are larger, but still 
a useful check against certain modified gravity models could be obtained. The full-scale experiment 
can offer good precision throughout the interested parameter ranges.


From the cosmographic measurments $D_A(z),H(z),f(z)$, one can constrain the 
cosmological model parameters. In this paper, we consider
a redshift-dependent equation of state for the dark energy parameterized
in the form of
\begin{equation}
\label{eqa:eos}
w(z)=w_0+w_a [1-a(z) ]=w_0+w_a\frac{z}{1+z}.
\end{equation}
The Fisher matrix of dark energy
parameters $w_0$ and $w_a$ is obtained by converting from the 
parameter space
$\{p_\alpha\} =\{D_{\rm A,i},~ H_i,~ f_i\}$ to the dark energy parameter space
$\{q_m\} =\{ w_0,~w_a,~\Omega_X \}$, using
\begin{equation}
F^{DE}_{mn}=\sum_{\alpha,\beta}\frac{\partial {\it p}_\alpha}{\partial {\it
    q}_m}F^{dis}_{\alpha \beta} \frac{\partial
 {\it p}_\beta}{\partial {\it q}_n}.
\end{equation}
To help break the parameter degeneracy between 
parameters, we combine the BAO data from Tianlai intensity mapping observation
with the data obtained from CMB observations.
The total Fisher matrix is given by \citep{2009MNRAS.394.1775W}
\begin{equation}
\label{eqa:Fadd}
F^{tot}_{\alpha \beta}= F^{CMB}_{\alpha \beta}+\sum_{i} F^{IM}_{\alpha \beta}(z_i),
\end{equation}
where $ F^{IM}_{\alpha \beta}(z_i)$ is the Fisher matrix derived from the $i$-th
redshift bin of the large scale structure imtensity mapping, and
$F^{CMB}_{\alpha \beta}$ is the CMB Fisher matrix.

\begin{figure}[t]
\centering{
\includegraphics[scale=0.35]{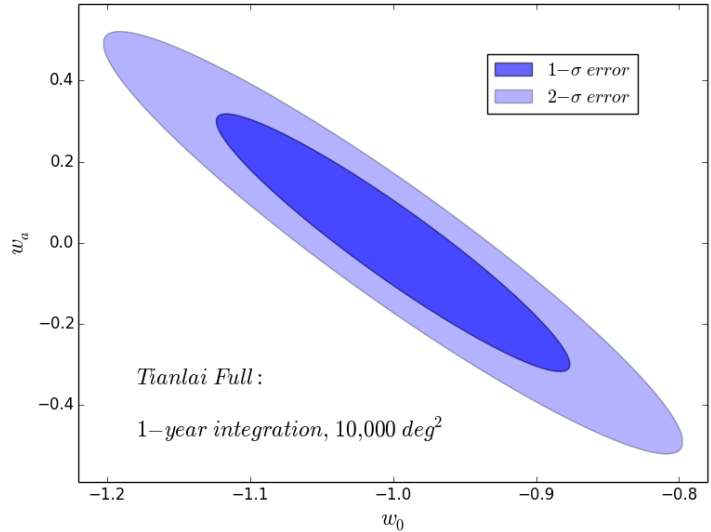}
\caption{Constraints on dark energy equation of state parameters $w_0$ and $w_a$
from full-scale Tianlai experiments.
The two contours are for 1-$\sigma$ and 2-$\sigma$ constraints respectively.
The integration time is 1 year, and the survey area is assumed to be 10000 $deg^2$.}
\label{Fig.DE_cont}
}
\end{figure}

The 1-$\sigma$ and 2-$\sigma$
measurement error contours for the variable dark energy equation of state parameters 
 $w_0$ and $w_a$ are shown in Fig.~\ref{Fig.DE_cont} for the
full-scale Tianlai experiment. Here we have assumed that the frequency range probed is 
400 -- 1420 MHz, the usable survey area is $10000\deg^2$, and the integration time is 1 year. The 
measurement error is obtained with a joint constraint with the CMB data, but no other 
observational data. We expect $\sigma_{w_0} \approx 0.0815$ and $\sigma_{w_a} \approx 0.210$. 
This is comparable with the precision of stage IV dark energy experiments as defined in
the Dark Energy Task Force (DETF) report \citep{2006astro.ph..9591A}.

\section{Fisher Forecasts for the Primordial Non-Gaussianity}

Typical single field slow roll
inflation models predict that the primordial density fluctuations follow the Gaussian
distribution, though the density distribution deviates from Gaussianity
as the structures grow and non-linearities appear. Detection or constraint on the primordial
non-Gaussianity will provide invaluable information on the origin of the Universe.

Compared with the observable galaxies which correspond to high density peaks of the
matter density distribution, the neutral hydrogen gas that exists in galaxies of
almost all mass scales is a less biased tracer of the underlying matter density, allowing the
primordial non-Gaussianity to be investigated from a different perspective.

The non-Gaussianity of the primordial density fluctuations can induce a
scale-dependent and redshift-dependent HI bias, similar to other biased tracers
\citep{2008PhRvD..77l3514D,2008ApJ...677L..77M}. This effect can be used to constrain the
primordial non-Gaussianity.
\citet{2013PhRvL.111q1302C} has demonstrated that a small but compact array
working at $\sim 400 \MHz$ would be
possible to place tight constraints on the $f_{\rm NL}$, with an error
close to $\sigma_{f_{\rm NL}} \sim 1$. We shall make forecast for making such constraints with
the Tianlai experiment.

Once the large scale structure of the 21 cm brightness temperature
fluctuations are mapped out, this same set of data can also be used to
measure the bispectrum of HI gas distribution. The HI bispectrum consists
contributions from primordial non-Gaussianity, the non-linear gravity evolution,
and the non-linear HI bias. The relative importance of primordial non-Gaussianity
increases toward higher redshifts \citep{2007PhRvD..76h3004S,2009ApJ...703.1230J}. The
21 cm experiment can in principle observe the 
large scale structure at relatively high redshifts from the ground without being 
affected significantly by the atomosphere, and this is  
an advantage of this method, though at present the 21 cm observation is still limited to
lower redshifts than the optical observations.
The 21 cm intensity mapping is much more efficient with large survey
volume without resolving individual galaxies.
Using the 21 cm bispectrum from the dark ages, \citet{2007ApJ...662....1P}
found that very low frequency
radio observations with high angular resolution could potentially detect
 primordial non-Gaussianity with $f_{\rm NL} \sim 1$.
Here we focus on the HI bispectrum after reionization, and assess the
constraining power of the 21 cm bispectrum
measured by Tianlai experiment.


\subsection{Constraints on $f_{\rm NL}$ from the HI Power Spectrum}\label{HI_PS}

The non-Gaussianity in the primordial density fluctuations can result in a scale-dependence
in the halo bias, which originates from coupling between large and small scales modes
\citep{2008PhRvD..77l3514D,2008ApJ...677L..77M}.
For the standard local type primordial non-Gaussianity,
the scale-dependent non-Gaussian correction to the linear halo bias, to the leading order, is
(see e.g. \citealt{2011PhRvD..84f3512D,2012PhRvD..86f3526A,2013MNRAS.428.2765D})
\begin{equation}
\Delta b^d(k,z) \;=\; \frac{2\, f_{\rm NL}\, (b_1^{\rm G} \,-\, 1)\, \delta_c}{{\mathcal M}(k,z)},
\end{equation}
where $b_1^{\rm G}$ is the linear halo bias for the Gaussian density field, $\delta_c = 1.686$ 
is the critical overdensity for spherical collapse, and ${\mathcal M(k,z)}$ relates the 
density fluctuations in Fourier space, $\delta_k$, to the primordial curvature 
perturbation, $\Phi_k$, via
the Poisson equation:
\begin{equation}
\delta_k(z) = {\mathcal M(k;z)} \Phi_k,
\end{equation}
where
\begin{equation}
{\mathcal M(k;z)} = \frac{2}{3}\, \frac{k^2\,T(k)\,G^+(z)\,c^2} {\Omega_{\rm m}\, H_0^2}.
\end{equation}
Here $T(k)$ is the matter transfer function normalized to unity on large scales, 
$c$ is the speed of light, and $G^+(z) = g(0)G(z)$ is the growth factor of the growing 
mode of density perturbations, in which $g(0) = (1+z_{\rm i})^{-1}G^{-1}(z_{\rm i})$ 
with $z_{\rm i}$ being the initial redshift, and $G(z)$ is the linear growth factor 
normalized to unity at $z=0$.
The corrected linear halo bias is $b_1 (k,M,z)= b_1^{\rm G}(M,z) + \Delta b^d(k,M,z)$.


The power spectrum of the density fluctuations of HI gas is
$P_{\rm HI}(k,z) = \left[b_1^{\rm HI}(k,z) \right]^2\,P_{\rm L}(k,z)$,
in which $P_{\rm L}(k,z)$ is the linear matter power spectrum, and the
scale-dependent HI bias $b_1^{\rm HI}(k,z)$ is related to the corrected linear halo bias
via the model described in section~\ref{P_signal}.
The observed power spectrum in redshift space after averaging over
angles in $k$ space is \citep{1997MNRAS.284..885P}
\begin{equation}
P_{\rm s} (k,z) = a_0^{\rm P}(\beta)\, P_{\rm HI}(k,z),
\end{equation}
where $a_0^{\rm P}(\beta) \;=\; 1 + \frac{2}{3}\beta + \frac{1}{5}\beta^2$,
with $\beta = \Omega_{\rm m}^{0.55}(z) / b_1^{\rm HI}$.
We apply the same Fisher matrix as Eq.~(\ref{Eq.fisher_PS}), but
here we take $f_{\rm NL}$ as the single parameter, and fix all the other cosmological parameters.


When adding information from all available wavenumbers in the Fisher matrix,
$k_{\rm max}$ is limited by the Nyquist frequency, $k_{\rm Nyq} = \pi / {\rm resolution}$,
which arises from the non-zero beam size of the cylinder array \citep{2010ApJ...721..164S},
as well as by the non-linear wavenumber cutoff, $k_{\rm nonl}$, above which
the linear power spectra are not accurate.
Here we adopt conservative values for $k_{\rm nonl}$ by
requiring $\sigma(R=\pi/2k_{\rm nonl}; z) = 0.5$
at each redshift bin \citep{2003ApJ...598..720S}.
Therefore, $k_{\rm max} = {\rm MIN} \{k_{\rm Nyq}, k_{\rm nonl}\}$.
Effectivley, $k_{\rm max}$ is limited by Nyquist wavenumber for pathfinder and
pathfinder+, while for the full-scale Tianlai, $k_{\rm max}$ is mostly set by
the non-linear cutoff, except for the highest
redshift bin. On the other hand, $k_{\rm min}$ is set by the scale defined
by the size of each redshift bin.

With the same survey parameters and redshift bins as in section \ref{fisher_DE},
we find that the constraint on the non-linear parameter $f_{\rm NL}$ for the local model
is quite weak for the pathfinder and pathfinder+ experiments. With the full-scale Tianlai experiment,
we can achieve $\sigma_{\rm f_{NL}}^{\rm local} \sim 14$.
The exact numbers of the predicted
$1-\sigma$ errors for Tianlai pathfinders and for the full-scale Tianlai are listed 
in table \ref{tab:PSerror1sigma}. 

\begin{table}[htb]
\caption{The predicted $1-\sigma$ errors of $f_{\rm NL}$ using HI power spectrum measured by Tianlai
\label{tab:PSerror1sigma}}
\begin{center}
\begin{tabular}{cccc}
\hline
\hline
\\
& pathfinder & pathfinder+ & full scale\\
\hline
$N_{\rm feed}$ per cylinder & 32 & 72 & 256\\
\hline
$\sigma_{\rm f_{NL}}^{\rm local}$   & 1504 & 161 & 14.1 \\
\hline
\end{tabular}
\end{center}
\end{table}

\subsection{Constraints on $f_{\rm NL}$ from the HI Bispectrum}\label{HIbispectrum}

On large scales, the matter bispectrum is well described by the tree-level expression, and the
loop corrections remain very small \citep{2013PhRvD..87d3512T,2014PhRvD..89b3516G}.
Higher order terms such as the trispectrum could contribute significantly
to the bispectrum of high density peaks \citep{2009PhRvD..80l3002S,2009ApJ...703.1230J}, but
as the HI gas is much less biased than observable galaxies -- for the Tianlai experiment, the HI
bias is not far from 1 -- we expect such contribution to be less significant, 
though the exact amount can not be obtained without going through the lengthy calculations.
Here we neglect the higher order terms
and account only for the tree level matter bispectrum here, and reserve
the investigation of the contribution from matter trispectrum to the HI
bispectrum to future works. If such contribution is significant, it would increase
the HI bispectrum, and we would obtain a stronger constraint on
$f_{\rm NL}$, so our current estimate may be regarded as a relatively conservative one.

As we are interested in forecasting the constraining power of HI bispectrum observations on the
primordial non-Gaussianity, i.e. the parameter $f_{\rm NL}$, in the following, we shall focus on
the reduced HI bispectrum, $Q_{\rm HI}$, which is much less sensitive to other
cosmological parameters \citep{2007PhRvD..76h3004S}.
In the real experiments, we always measure the 21 cm brightness temperature in redshift space.
Similar to the tree-level expresion for the observed galaxy bispectrum \citep{2007PhRvD..76h3004S},
the reduced HI bispectrum
in redshift space after averaging over angles in $k$ space is
\begin{equation}
Q_s (k_1,k_2,k_3) \;=\; \frac{a_0^{\rm B}(\beta)}{\left[a_0^{\rm P}(\beta)\right]^2}\, 
\left[\frac{1}{b_1^{\rm HI}}\, Q^{\rm tree} (k_1,k_2,k_3) \,+\, \frac{b_2^{\rm HI}}{(b_1^{\rm HI})^2}\right],
\end{equation}
where $a_0^{\rm B}(\beta) \;=\; 1 + \frac{2}{3}\beta + \frac{1}{9}\beta^2$ converts the bispectrum
from real space to redshift space, and $Q^{\rm tree}$ is the reduced tree-level bispectrum of underlying matter.
The first term includes the contributions from primordial non-Gaussianity and non-linear
gravitational evolution, and the second represents the contribution from non-linear bias of HI gas.



The reduced matter bispectrum can be written as the sum of two contibutions:
\begin{eqnarray}\label{Qtree}
Q^{\rm tree} (k_1,k_2,k_3) &=& Q_{\rm I} (k_1,k_2,k_3) \,+\, Q_{\rm G} (k_1,k_2,k_3) \nonumber
\\
&=& \frac{B_{\rm I} (k_1,k_2,k_3)}{P_{\rm L}(k_1)\,P_{\rm L}(k_2) \,+\, (2\, perm.)} \,+\, \frac{B_{\rm G} (k_1,k_2,k_3)}{P_{\rm L}(k_1)\,P_{\rm L}(k_2) \,+\, (2\, perm.)},
\end{eqnarray}
where $+\, (n\, perm.)$ stands for the sum of $n$ additional terms
permuting $k_1$, $k_2$, and $k_3$.
The matter bispectrum due to gravity alone, $B_{\rm G}$, is given by the
second order perturbation theory
\citep{1984ApJ...279..499F,2002PhR...367....1B}, and the
the matter bispectrum contributed from primordial non-Gaussianity, $B_{\rm I}$, is related to
the bispectrum of curvature perturbations, $B_{\rm \Phi}$, by
\begin{equation}\label{Eq.BI}
B_{\rm I} (k_1,k_2,k_3) \;=\; {\mathcal M}(k_1;z)\,{\mathcal M}(k_2;z)\,{\mathcal M}(k_3;z)\, B_{\rm \Phi} (k_1,k_2,k_3).
\end{equation}

We consider two models of primordial non-Gaussianity here, i.e.
the local model and the equilateral model,
but the same forecast can also be applied to other models of interest.
The local model is physically motivated, in this case the contributions from the
squeezed triangular configurations dominate.
The leading contribution to the $f_{\rm NL}^{\rm local}$ expansion of the bispectrum of
curvature perturbation is
\begin{eqnarray}\label{Eq.Bphi_local}
B_{\rm \Phi}^{\rm local}(k_1,k_2,k_3) &\simeq& 2\,f_{\rm NL}^{\rm local}\, [P_{\rm \Phi}(k_1)P_{\rm \Phi}(k_2) + P_{\rm \Phi}(k_2)P_{\rm \Phi}(k_3) + P_{\rm \Phi}(k_3)P_{\rm \Phi}(k_1)] \nonumber
\\
&=& 2\,f_{\rm NL}^{\rm local}\, \Delta_{\rm \Phi}^2\, \left[ \frac{1}{k_1^{4-n_s}\,k_2^{4-n_s}} \,+\, (2\, perm.)\right],
\end{eqnarray}
where $\Delta_{\rm \Phi} \equiv P_{\rm \Phi}/k^{n_s-4}$, and $ P_{\rm \Phi}(k)$ is
the curvature power spectrum.
The equilateral model is a good approximation to the higher derivative
models \citep{2003JCAP...10..003C} and the DBI inflationary
model\citep{2004PhRvD..70l3505A}. The bispectrum of curvature perturbation
for the equilateral model is
\begin{eqnarray}\label{Eq.Bphi_eq}
B_{\rm \Phi}^{\rm equil.} =& 6\,f_{\rm NL}^{\rm equil.}\, \Delta_{\rm \Phi}^2\, \left[\,-\,\frac{1}{k_1^{4-n_s}\, k_2^{4-n_s}} \,-\, \frac{1}{k_2^{4-n_s}\, k_3^{4-n_s}} \,-\, \frac{1}{k_3^{4-n_s}\, k_1^{4-n_s}} \,-\, \frac{2}{(k_1\, k_2\, k_3)^{2(4-n_s)/3}} \right.\nonumber
\\
&+\left. \left(\frac{1}{k_1^{(4-n_s)/3}\, k_2^{2(4-n_s)/3}\, k_3^{4-n_s}} \,+\, (5\, perm.) \right) \,\right].
\end{eqnarray}



Following \citet{1998ApJ...496..586S}, a bispectrum estimator for a cubic survey volume of $V$ can be defined as
\begin{equation}
\hat{B}(k_1,k_2,k_3)\;\equiv\; \frac{V_{\rm f}}{V_{\rm B}(k_1,k_2,k_3)}\, \int_{k_1} d^3\vec{q}_1\, \int_{k_2} d^3\vec{q}_2\, \int_{k_3} d^3\vec{q}_3\, \delta_{\rm D}(\vec{q}_1+\vec{q}_2+\vec{q}_3)\, \delta(\vec{q}_1)\, \delta(\vec{q}_2)\, \delta(\vec{q}_3),
\end{equation}
where $V_{\rm f} \equiv k_{\rm f}^3 = (2\pi)^3/V$ is the elemental volume in $k$ space of observation cells, and each integration is over the range $[k_i-\Delta k/2, k_i+\Delta k/2]$ centered on $k_i$, with $\Delta k$ equal to a multiple of $k_{\rm f}$. Here $\delta_{\rm D}(\vec{q}_1+\vec{q}_2+\vec{q}_3)$ is the Dirac delta function which ensures that the vectors $\vec{q}_1$, $\vec{q}_2$, and $\vec{q}_3$ form a triangle, while $V_{\rm B}(k_1,k_2,k_3)$ is the normalization factor given by
\begin{equation}
V_{\rm B} \;\equiv\; \int_{k_1} d^3\vec{q}_1\, \int_{k_2} d^3\vec{q}_2\, \int_{k_3} d^3\vec{q}_3\, \delta_{\rm D}(\vec{q}_1+\vec{q}_2+\vec{q}_3) \;\simeq\; 8\pi^2\, k_1\,k_2\,k_3\, \Delta k_1\, \Delta k_2\, \Delta k_3.
\end{equation}
In the following, we assume $\Delta k_i = k_{\rm f}$ so as to take into account all ``fundamental'' triangular configurations.

To the leading order, the variance of this estimator is \citep{1998ApJ...496..586S}
\begin{equation}\label{DeltaB2}
\Delta B_s^2(k_1,k_2,k_3) \;\simeq\; (2\pi)^3\, V_{\rm f}\, \frac{s_{\rm 123}}{V_{\rm B}}\, P_{\rm tot}(k_1)\, P_{\rm tot}(k_2)\, P_{\rm tot}(k_3),
\end{equation}
where $s_{\rm 123} = 6, 2, 1$ respectively for equilateral, isosceles and general triangles, and $P_{\rm tot}(k)$ is total measured power spectrum including the redshift space HI power spectrum, $P_{\rm s}(k) = a_0^{\rm P}(\beta)\,P_{\rm HI}(k)$, and the noise power spectrum $N(k)$.

The Fisher matrix for observations of reduced bispectrum at a given redshift bin can be written as
\begin{equation}
F_{\alpha \beta} \;\equiv\; \sum_{k_1=k_{\rm min}}^{k_{\rm max}}\, \sum_{k_2=k_{\rm min}}^{k_1}\, \sum_{k_3=k_{\rm min}^\star}^{k_2}\, \frac{\partial Q_s}{\partial \alpha}\, \frac{\partial Q_s}{\partial \beta}\, \frac{1}{\Delta Q_s^2},
\end{equation}
where $\Delta Q_s^2$ is the variance of the reduced HI bispectrum measured in redshift space, and
$\alpha$ and $\beta$ represent the parameters we are interested in, i.e. $f_{\rm NL}$, and
$b_1^{\rm HI} (z_i)$ and $b_2^{\rm HI}(z_i)$ for each redshift bin $z_i$ of the survey.
The three sums are over all combinations of $k_1$, $k_2$ and $k_3$ that form triangles,
in steps of $\Delta k_i$, with $k_{\rm min}^\star = \max (k_{\rm min}, |k_1-k_2|)$.
In each redshift bin, we divide the survey volume into cubes, and $k_{\rm min}$ is still set by
the scale spanning the redshift bin.
$k_{\rm max}$ is set by the Nyquist frequency or the smallest scale at which we can trust
our model for the HI bispectrum. Here we assume the tree-level bispectrum breaks down
below the non-linear scale cutoff, so that $k_{\rm max} = {\rm Min} \{k_{\rm Nyq}, k_{\rm nonl}\}$.
If we assume the variance of the HI bispectrum $\Delta B_s$ dominates over the
variance of the HI power spectrum $\Delta P_s$, 
then the variance of the reduced HI bispectrum
in redshift space can be written as \citep{2007PhRvD..76h3004S}:
\begin{equation}
\Delta Q_s^2 (k_1,k_2,k_3) \;\simeq\; \frac{\Delta B_s^2(k_1,k_2,k_3)}{\left[P_s(k_1)P_s(k_2) \,+\, (2\,perm.)\right]^2},
\end{equation}
with the $\Delta B_s^2$ given by Eq.~(\ref{DeltaB2}).

We assume the the fiducial values of HI bias parameters as given in section \ref{P_signal},
and take the fiducial value of $f_{\rm NL}=0$ for both the local and equilateral models. 
Assuming 1 year's integration time and a total survey area of 10000 deg$^2$,
the marginalized $1-\sigma$ errors on $f_{\rm NL}^{\rm local}$ and $f_{\rm NL}^{\rm equil}$
are listed in table \ref{tab:BSerror1sigma}.
Again, we find that the pathfinder and pathfinder+ data are insufficient to 
 provide much constraint on the bispectrum, due to the large error in its measurement.
With the full-scale Tianlai experiment, we could achieve
$\sigma_{\rm f_{NL}}^{\rm local} \sim 22$ for the local model,
and $\sigma_{\rm f_{NL}}^{\rm equil} \sim 157$ for the equilateral model.

\begin{deluxetable}{cccccc}
\tablecaption{The marginalized $1-\sigma$ errors of $f_{\rm NL}$ using HI bispectrum measured by Tianlai 
\label{tab:BSerror1sigma}}
\tablewidth{0pt}
\tablehead{
& \colhead{Pathfinder}& \colhead{Pathfinder+}& \colhead{Full scale}\\
\colhead{$N_{\rm feed}$ per cylinder} &\colhead{32}&\colhead{72}&\colhead{256}
}
\startdata
$\sigma_{\rm f_{NL}}^{\rm local}$            & 70814 & 2272 & 21.7 \\
$\sigma_{\rm f_{NL}}^{\rm equil}$            & 79427 & 2754 & 157 \\
\enddata
\end{deluxetable}


\section{Conclusions}\label{conclusions}

In this work, we assess the capability of the Tianlai experiments in constraining
various cosmological parameters, specifically the dark energy equation of state
and the level of primordial non-Gaussianity. We use the Fisher information matrix 
method, which are widely used for making such forecasts.
We have compared our results with other forecasts on 21 cm intensity 
mapping experiments \citep{2008PhRvL.100i1303C,2008arXiv0807.3614A,2010ApJ...721..164S,
2012A&A...540A.129A,2014MNRAS.444.3183A}, and found they generally yield 
similar results when the same
conditions are assumed.

Currently, our plan is to first test the principle and key technologies 
with a smaller scale pathfinder experiment, then 
upgrade to the pathfinder+ experiment, before eventually building the 
the full-scale Tianlai experiment. 
The goal of the pathfinders is to test the technologies and feasibility
of HI intensity mapping observations with cylinder arrays, and as shown in this work,
we expect to be able to measure the HI power spectrum with the pathfinders, but
the constraints obtainable on cosmological parameters would be fairly weak.

The full-scale Tianlai experiment will significantly tighten the constraints by adding the
number of receivers, hence increasing the effective collecting area of the cylinders,
and by expanding the scale of the cylinders, hence increasing the spatial resolution.
Assuming an integration time of 1 year, and a survey area of 10000 $\deg^2$,
we expect $\sigma_{w_0} \sim 0.082$ and $\sigma_{w_a} \sim 0.21$
from the BAO and RSD measurements. This is comparable to the expected precision from
stage IV dark energy experiments as defined by the DETF report \citep{2006astro.ph..9591A}, 
while the cost would only be a small fraction of the such experiments.

The primordial non-Gaussianity can be constrained by looking for scale-dependent bias of the 
power spectrum, or by bispectrum measurement. We find
$\sigma_{\rm f_{NL}}^{\rm local} \sim 14$ from
the power spectrum measurements with scale-dependent bias, and
$\sigma_{\rm f_{NL}}^{\rm local} \sim 22$ and $\sigma_{\rm f_{NL}}^{\rm equil} \sim 157$
from the bispectrum measurements. The constraints on the primordial non-Gaussianity 
from large scale structure observations including this one
are generally weaker than from high precision CMB observations, but they probe different
scales, so it is still very important to do such observations. 

In making these forecasts, we have assumed that the foregrounds can be effectively removed, 
and their residuals only affect the overal system temperature. 
In fact, due to the complicated system responses
and our imperfect knowledge of it from calibration, the foreground removal will 
not be so effective, and the foreground may also introduce non-Gaussian features which 
could potentially contaminate the measurement of the primordial non-Gaussianity. These 
problems may degrade the measurement precision \citep{2012A&A...540A.129A}. Therefore, the results
presented here should be regarded as an ideal case. We plan to make more detailed simulations
to assese the effects of calibration and foreground subtraction on the measurement process for 
the Tianlai experiments in the near future. 










\acknowledgments
We thank Fengquan Wu, Yi Mao, Hong Guo, Junqing Xia, and Kwan Chuen Chan for many
helpful discussions. This work is supported by the MoST 863
program grant 2012AA121701, the NSFC grants 11373030 and 11303034, and 
the CAS Strategic Priority Research Program ``The Emergence of
Cosmological Structures" Grant No. XDB09000000.

\bibliography{references}
\bibliographystyle{apj}






\end{document}